\documentclass[a4paper,12pt]{article}
\usepackage{amssymb}
\usepackage{amsmath}
\usepackage{color}
\usepackage[unicode,bookmarks,bookmarksopen,bookmarksopenlevel=2,colorlinks,linkcolor=blue,citecolor=green]{hyperref}

\input{tcilatex}
\begin{document}

\title{\textbf{The Constant Astigmatism Equation. New Exact Solution.}}
\author{Natale Manganaro$^{1}$, Maxim V.~Pavlov$^{2,3,4}$ \\
%EndAName
$^{1}$Department of Mathematics and Informatics,\\
University of Messina,\\
Viale Ferdinando Stango D'Alcontres 31, 98166 Messina, Italy\\
$^{2}$Sector of Mathematical Physics,\\
Lebedev Physical Institute of Russian Academy of Sciences,\\
Leninskij Prospekt 53, 119991 Moscow, Russia\\
$^{3}$Laboratory of Geometric Methods in Mathematical Physics,\\
Lomonosov Moscow State University,\\
Leninskie Gory 1, 119991, Moscow, Russia\\
$^{4}$Mathematical institute in Opava,\\
Silesian University in Opava,\\
Na Rybn\'{\i}\v{c}ku 1, 746 01 Opava, Czech Republic }
\date{}
\maketitle

\begin{abstract}
In this paper we present a new solution for the Constant Astigmatism
equation. This solution is parameterized by an arbitrary function of a
single variable.
\end{abstract}

\tableofcontents

\textit{keywords}: Constant Astigmatism equation, differential constraints,
overdetermined systems, characteristic method, integrable systems,
reductions.

\bigskip



\section{Introduction}

The theory of integrable systems usually associates with the method of the
Inverse Scattering Transform, which allows to construct multi-soliton,
multi-gap and similarity solutions, i.e. parameterized by an arbitrary
number of constants. Nevertheless, some integrable nonlinear systems in
partial derivatives possess solutions parameterized by arbitrary functions
of a single variable. The most known example is the three-wave interaction,
which were deeply investigated in \cite{ZM}.

In this paper we deal with the Constant Astigmatism equation%
\begin{equation}
u_{tt}+\left( \frac{1}{u}\right) _{xx}+2=0,  \label{asti}
\end{equation}%
which was considered in a set of papers (see detail in \cite{MM0}, \cite{MM}%
, \cite{MM1}, \cite{MM2}, \cite{MM3}, \cite{PZ}). This equation is connected
with the remarkable Bonnet (also known as the Sine-Gordon) equation by a
reciprocal transformation. However, we show that just the Constant
Astigmatism equation has the particular solution parameterized by an
arbitrary function of a single variable.

To illustrate this phenomenon, we just mention another remarkable integrable
nonlinear system known as the Kaup--Boussinesq system%
\begin{equation}
u_{t}+\left( \frac{u^{2}}{2}-\eta \right) _{x}=0,\text{ \ \ }\eta
_{t}+\left( u\eta -\frac{1}{4}u_{xx}\right) _{x}=0.  \label{kb}
\end{equation}%
Under the potential substitution $u=z_{x}$ and $\eta =z_{x}^{2}/2+z_{t}$,
the second equation becomes%
\begin{equation*}
z_{tt}+2z_{x}z_{xt}+\left( z_{t}+\frac{3}{2}z_{x}^{2}\right) z_{xx}-\frac{1}{%
4}z_{xxxx}=0.
\end{equation*}%
This equation possesses the reduction%
\begin{equation*}
z_{t}+\frac{1}{2}(z_{x}^{2}-z_{xx})=0,
\end{equation*}%
which is nothing but a well-known Burgers equation%
\begin{equation}
u_{t}+uu_{x}-\frac{1}{2}u_{xx}=0,  \label{bur}
\end{equation}%
which is linearizable to the heat equation%
\begin{equation*}
v_{t}=\frac{1}{2}v_{xx},
\end{equation*}%
by the Cole--Hopf substitution $u=-v_{x}/v$ (indeed, one can see that the
first equation in (\ref{kb}) reduces to (\ref{bur}) if $\eta =u_{x}/2$; then
the second equation in (\ref{kb}) automatically satisfies). Since, the heat
equation has a solution parameterized by an arbitrary function of a single
variable $v_{0}(x)$, then the Kaup--Boussinesq system also admits a
particular solution $u=z_{x}$ and $\eta =z_{x}^{2}/2+z_{t}$, where:%
\begin{equation*}
u=-\partial _{x}\ln \left[ \frac{1}{\sqrt{2\pi t}}\underset{-\infty }{%
\overset{\infty }{\int }}\exp \left( -\frac{(x-y)^{2}}{2t}-\underset{0}{%
\overset{y}{\int }}u(\xi ,0)d\xi \right) dy\right] ,
\end{equation*}%
\begin{equation*}
\eta =-\frac{1}{2}\partial _{x}^{2}\ln \left[ \frac{1}{\sqrt{2\pi t}}%
\underset{-\infty }{\overset{\infty }{\int }}\exp \left( -\frac{(x-y)^{2}}{2t%
}-\underset{0}{\overset{x^{\prime }}{\int }}u(\xi ,0)d\xi \right) dy\right] ,
\end{equation*}%
which depends on an arbitrary function $u_{0}(x)=u(x,t)|_{t=0}$.

\section{Reduction Procedure}

The aim of this section is to develop a reduction procedure for (\ref{asti})
within the framework of the theory of differential constraints. The method
is based upon appending a set of PDEs to a given governing system of field
equations and it was first applied by Janenko \cite{jan} to the gas dynamics
model. The auxiliary equations play the role of differential constraints
because they select classes of solutions of the system under interest. The
method is very general and, in fact, it includes many of the reduction
approaches known in a literature. Unfortunately often the generality of the
method limits according diversity of applications. In each particular case,
the method of differential constraints utilizes specific features of a
corresponding nonlinear system (see Refs. \cite{man4}, \cite{mel1}, \cite%
{kap}, \cite{cur1}, \cite{cur2}, \cite{cur3}, \cite{cur4}, \cite{cur5}).

First, for further convenience we change the sign $u\rightarrow -u$ in the
Constant Astigmatism equation (\ref{asti})%
\begin{equation}
u_{tt}=2-\left( \frac{1}{u}\right) _{xx}.  \label{complex}
\end{equation}%
Next, we reduce the governing equation (\ref{complex}) to the pair of
equations%
\begin{equation}
\left\{ 
\begin{array}{l}
a_{t}-b_{x}=0 \\ 
\\ 
b_{t}-\frac{1}{\left( a+t^{2}\right) ^{2}}a_{x}=0%
\end{array}%
\right.   \label{sis}
\end{equation}%
where%
\begin{equation}
a(x,t)=u(x,t)-t^{2},  \label{transf}
\end{equation}%
and $b(x,t)$ is an auxiliary function. System (\ref{sis}) is strictly
hyperbolic and the eigenvalues of the matrix coefficients as well as the
corresponding left eigenvectors are, respectively,%
\begin{equation}
\lambda =\mp f(a,t);\quad \quad \mathbf{l}^{\left( \mp \right) }=\left(
-\lambda ;\;\;1\right) ,  \label{eigen}
\end{equation}%
where%
\begin{equation}
f(a,t)=\frac{1}{a+t^{2}}.  \label{f}
\end{equation}%
Therefore, according to \cite{mel2}, owing to (\ref{eigen}), the more
general first order differential constraint admitted by (\ref{sis}) takes
the form%
\begin{equation*}
\mathbf{l}^{\left( \mp \right) }\cdot \mathbf{U}_{x}=p(x,t,a,b),\quad %
\mbox{with}\quad \mathbf{U}=\left( 
\begin{array}{l}
a \\ 
b%
\end{array}%
\right) ,
\end{equation*}%
which in our case specializes to%
\begin{equation}
b_{x}-\lambda a_{x}=p(x,t,a,b),  \label{const}
\end{equation}%
where the function $p(x,t,a,b)$ must be determined during the process. In
the following without loss of generality we consider the case $\lambda
=+f(a,t)$. The consistency requirement between (\ref{sis}) and (\ref{const})
leads to a linear expression with respect to $a_{x}$ whose coefficients
depend only on $(x,t,a,b)$. Thus both coefficients must vanish independently:%
\begin{equation}
\left\{ 
\begin{array}{l}
p_{a}+fp_{b}=-\frac{f_{t}+pf_{a}}{2f} \\ 
\\ 
p_{t}+fp_{x}=\frac{f_{t}+pf_{a}}{2f}p.%
\end{array}%
\right.   \label{comp}
\end{equation}%
Both these equations can be integrated by the method of characteristics,
i.e. we have%
\begin{equation*}
\frac{db}{da}=f,\text{ \ }\frac{dp}{da}=-\frac{f_{t}+pf_{a}}{2f},\text{ \ }%
\frac{dx}{dt}=f,\text{ \ }\frac{dp}{dt}=\frac{f_{t}+pf_{a}}{2f}p.
\end{equation*}%
Taking (\ref{f}) into account, we can independently solve first two ordinary
differential equations (with respect to $a$) and two other ordinary
differential equations (with respect to $t$). However, the solution of (\ref%
{comp})%
\begin{equation}
p=-2t\pm 2\sqrt{a+t^{2}}  \label{p}
\end{equation}%
follows directly from the compatibility condition%
\begin{equation*}
\frac{d}{dt}\left( \frac{dp}{da}\right) =\frac{d}{da}\left( \frac{dp}{dt}%
\right) .
\end{equation*}%
Therefore, taking (\ref{const}) and (\ref{p}) into account, equations (\ref%
{sis}) can be written in the form%
\begin{equation}
\left\{ 
\begin{array}{l}
a_{t}-fa_{x}=p \\ 
\\ 
b_{t}-fb_{x}=-fp.%
\end{array}%
\right.   \label{sis1}
\end{equation}%
Then the first equation%
\begin{equation}
a_{t}-\frac{1}{a+t^{2}}a_{x}=-2t\pm 2\sqrt{a+t^{2}}  \label{aa}
\end{equation}%
again can be integrated by the method of characteristic, while the function $%
b(x,t)$ satisfying the second equation can be found in quadratures (see (\ref%
{sis}))%
\begin{equation*}
db=a_{t}dx+\frac{a_{x}}{(a+t^{2})^{2}}dt\equiv (u_{t}-2t)dx+\frac{u_{x}}{%
u^{2}}dt,
\end{equation*}%
because the compatibility condition $(b_{t})_{x}=(b_{x})_{t}$ is fulfilled
by virtue of (\ref{complex}). Moreover, since $b_{t}=f^{2}a_{x}$ and $%
b_{x}=a_{t}$, one can see that the second equation in (\ref{sis1}) is
equivalent to the first one in (\ref{sis1}).

Of course a similar analysis holds in the case $\lambda =-f(a,t)$.

In order to obtain the required reduction of the governing equation (\ref%
{complex}), by writing equations (\ref{const}) and (\ref{sis1}) in terms of
the original variable $u$ we get (see (\ref{transf}) and, for instance, (\ref%
{aa}) for the case $\lambda =+f(a,t)$)%
\begin{equation}
u_{t}=\pm \frac{1}{u}u_{x}\pm 2\sqrt{u}.  \label{equa}
\end{equation}%
So that exact solutions of (\ref{complex}) are determined by integrating the
first order equation (\ref{equa}) using the characteristic method. Therefore
we proved the following:

\textbf{Theorem 1}: \textit{Constant Astigmatism equation} (\ref{complex}) 
\textit{possesses four natural reductions }(\ref{equa}). \textit{Thus,
Constant Astigmatism equation} (\ref{complex}) \textit{has four particular
solutions parameterized by an arbitrary function of a single variable.}

\section{New Particular Solution}

Let us first prove the following

\textbf{Theorem 2}: \textit{The first order equation}%
\begin{equation}
u_{t}=F(x,t,u,u_{x}).  \label{a}
\end{equation}%
\textit{is a reduction of Constant Astigmatism equation} (\ref{complex}) 
\textit{if it specializes to} (\ref{equa}).

\textbf{Proof}: Substitution of (\ref{a}) into (\ref{complex}) leads to%
\begin{equation}
F_{t}+FF_{u}+F_{y}\left( F_{x}+F_{u}u_{x}+F_{y}u_{xx}\right) -\frac{1}{u^{2}}%
u_{xx}+\frac{2}{u^{3}}u_{x}^{2}=2,  \label{F}
\end{equation}%
where we set $y=u_{x}$. By requiring (\ref{F}) is satisfied $\forall u_{xx}$
we get%
\begin{equation}
F=\pm \frac{y}{u}+G(x,t,u),  \label{b}
\end{equation}%
where the function $G(x,t,u)$ needs to be determined. Substitution of (\ref%
{b}) in (\ref{F}) yields%
\begin{equation}
G_{t}+\left( G\pm \frac{y}{u}\right) \left( G_{u}\mp \frac{y}{u^{2}}\right)
\pm \frac{1}{u}\left( G_{x}+\left( G_{u}\mp \frac{y}{u^{2}}\right) y\right) +%
\frac{2}{u^{3}}y^{2}=2.  \label{G}
\end{equation}%
Since relation (\ref{G}) must be satisfied $\forall y$, we obtain $G=\pm 2%
\sqrt{u}$. Thus, in virtue of (\ref{b}), the Theorem is proved.

Next we solve equation (\ref{equa}) by a slightly modified version of the
characteristic method.

Under the point transformation $u=v^{2}$ (\ref{equa}) reduces to the form%
\begin{equation*}
v_{t}=\pm \frac{1}{v^{2}}v_{x}\pm 1.
\end{equation*}%
This equation can be written in the conservative form%
\begin{equation*}
w_{t}\pm \left( \frac{1}{w\pm t}\right) _{x}=0,
\end{equation*}%
where $v=w\pm t$. Then the potential function $z$ can be introduced such that%
\begin{equation*}
dz=wdx\mp \frac{dt}{w\pm t}.
\end{equation*}%
In the first case%
\begin{equation*}
d[z+\ln (w+t)]=wdx+\frac{dw}{w+t}.
\end{equation*}%
The compatibility condition implies $\left( \frac{1}{w+t}\right) _{x}=1$.
Thus $\frac{1}{w+t}=x+h(w)$, where $h(w)$ is an arbitrary function. So, the
first particular solution of the Constant Astigmatism equation is%
\begin{equation}
u_{(1)}=(w+t)^{2},  \label{raz}
\end{equation}%
where $w(x,t)$ is a solution of the algebraic equation%
\begin{equation}
(x+h_{1}(w))(t+w)=1.  \label{dva}
\end{equation}%
Analogously, in the second case:%
\begin{equation*}
u_{(2)}=(w-t)^{2},
\end{equation*}%
where $w(x,t)$ is a solution of the algebraic equation%
\begin{equation*}
(x+h_{2}(w))(t-w)=1;
\end{equation*}%
In the third case:%
\begin{equation*}
u_{(3)}=(w+t)^{2},
\end{equation*}%
where $w(x,t)$ is a solution of the algebraic equation%
\begin{equation*}
(h_{3}(w)-x)(t+w)=1;
\end{equation*}%
In the fourth case:%
\begin{equation*}
u_{(4)}=(w-t)^{2},
\end{equation*}%
where $w(x,t)$ is a solution of the algebraic equation%
\begin{equation*}
(h_{4}(w)+x)(w-t)=1.
\end{equation*}

\textbf{Remark 1}: The Constant Astigmatism equation preserves itself under
the transformation $x\leftrightarrow t$ and $u\rightarrow 1/u$. Also its
reduction (\ref{equa}) preserves itself under the same transformation.

\textbf{Remark 2}: A relationship between the Sine-Gordon and the Constant
Astigmatism equations was presented in \cite{MM3}. Corresponding reciprocal
transformation (see formulas (29), (30) in this cited paper) contains two
distinct expressions%
\begin{equation*}
\left( \frac{u_{x}}{u}\pm u_{t}\right) ^{2}-4u,
\end{equation*}%
which \textit{vanishes} if the function $u(x,t)$ satisfies (\ref{equa}).
Thus, the above four particular solutions parameterized by an arbitrary
function of a single variable cannot be transformed to corresponding
solutions of the Sine-Gordon equation.

\textbf{Remark 3}: A Cauchy problem for a nonlinear equation in partial
derivatives of a second order is based on $u|_{t=0}=u_{0}(x)$ and $%
u_{t}|_{t=0}=u_{1}(x)$. Since, we can consider a nonlinear equation in
partial derivatives of a first order, we investigate a Cauchy problem
restricted on a narrow class of solutions. Indeed, for instance, in the
first above case, a Cauchy problem has a solution%
\begin{equation}
u=(W(x,t)+t)^{2},  \label{u}
\end{equation}%
where $W(x,t)$ is a solution of an algebraic equation%
\begin{equation*}
\left( x+\frac{1}{W}-X(W)\right) (t+W)=1,
\end{equation*}%
and the function $X(W)$ is determined by the equation $X(W_{0}(x))=x$, where 
$W_{0}(x)=\sqrt{u_{0}(x)}$ (see equation (\ref{u}) for $t=0$). Then $%
u_{1}(x) $ cannot be an arbitrary function:%
\begin{equation*}
u_{1}(x)=2\frac{1+X^{\prime }(W)W^{2}}{X^{\prime }(W)W}|_{W=\sqrt{u_{0}(x)}}.
\end{equation*}

\section{Replication of Solutions}

According to \cite{MM1}, particular solutions of the Constant Astigmatism
equation can be replicated infinitely many times starting from any initial
solution by the formulas%
\begin{equation}
x^{(1)}=\frac{xu}{x^{2}u-1},\text{ \ }t^{(1)}=\eta ,\text{ \ }u^{(1)}=\frac{%
(x^{2}u-1)^{2}}{u};  \label{tri}
\end{equation}%
\begin{equation}
x^{(-1)}=\xi ,\text{ \ }t^{(-1)}=\frac{t}{t^{2}-u},\text{ \ }u^{(-1)}=\frac{u%
}{(t^{2}-u)^{2}},  \label{tre}
\end{equation}%
where%
\begin{equation}
-d\eta =xu_{t}dx+\left( x\frac{u_{x}}{u^{2}}+\frac{1}{u}+x^{2}\right) dt,%
\text{ \ }d\xi =(tu_{t}-u-t^{2})dx+t\frac{u_{x}}{u^{2}}dt.  \label{five}
\end{equation}

Combination of these two transformations gives an infinite series of
particular solutions in a general case. In this Section we consider
replication of particular solutions starting from the solution constructed
in the previous Section. Without loss of generality, we consider the first
such a particular solution determined by (\ref{raz}), (\ref{dva}), i.e.%
\begin{equation*}
u=(w+t)^{2},
\end{equation*}%
where $w(x,t)$ is a solution of the algebraic equation%
\begin{equation}
x=\frac{1}{t+w}-h(w).  \label{quadro}
\end{equation}%
Then the functions $\eta ,\xi $ (see (\ref{five})) can be found in
quadratures, i.e.%
\begin{eqnarray}
&&t^{(1)}\equiv \eta =2h(w)-wh^{2}(w)+\int h^{2}(w)dw-h^{2}(w)t  \notag \\
&&x^{(-1)}\equiv \xi =-\frac{w^{2}}{t+w}+2w+\int w^{2}h^{\prime }(w)dw, 
\notag
\end{eqnarray}%
while (\ref{tri}) leads to a first iteration%
\begin{equation*}
u^{(1)}=(t^{(1)}+w^{(1)})^{2},
\end{equation*}%
where $w^{(1)}(x^{(1)},t^{(1)})$ is a solution of the algebraic equation
(cf. (\ref{quadro}))%
\begin{equation}
x^{(1)}=\frac{1}{t^{(1)}+w^{(1)}}-h^{(1)}(w^{(1)}).  \label{six}
\end{equation}%
In this case%
\begin{equation*}
t\equiv \eta ^{(1)}=2h^{(1)}(w^{(1)})-w^{(1)}[h^{(1)}(w^{(1)})]^{2}+\int
[h^{(1)}(w^{(1)})]^{2}dw^{(1)}-[h^{(1)}(w^{(1)})]^{2}t^{(1)},
\end{equation*}%
where%
\begin{equation*}
w^{(1)}=-\int h^{2}(w)dw,\text{ \ \ }h^{(1)}(w^{(1)})=\frac{1}{h(w)},\text{
\ }w=-\int [h^{(1)}(w^{(1)})]^{2}dw^{(1)}.
\end{equation*}%
Meanwhile, transformation (\ref{tre}) leads to a first \textquotedblleft
negative\textquotedblright\ iteration%
\begin{equation*}
u^{(-1)}=(w^{(-1)}+t^{(-1)})^{2},
\end{equation*}%
where $w^{(-1)}(x^{(-1)},t^{(-1)})$ is a solution of the algebraic equation
(cf. (\ref{quadro}) and (\ref{six}))%
\begin{equation*}
x^{(-1)}=\frac{1}{t^{(-1)}+w^{(-1)}}-h^{(-1)}(w^{(-1)}).
\end{equation*}%
In this case%
\begin{equation*}
t^{(-1)}=-\frac{t}{w^{2}+2wt},\text{ \ }t=-\frac{t^{(-1)}}{%
(w^{(-1)})^{2}+2w^{(-1)}t^{(-1)}},
\end{equation*}%
where%
\begin{equation*}
h^{(-1)}(w^{(-1)})=-\int w^{2}h^{\prime }(w)dw,\text{ \ }w^{(-1)}=w^{-1}.
\end{equation*}

Thus, we see that both transformations $(x,t,u)\rightarrow
(x^{(1)},t^{(1)},u^{(1)})$ and $(x,t,u)\rightarrow
(x^{(-1)},t^{(-1)},u^{(-1)})$ preserve the class of solutions (see Remark 2
in the previous Section) which cannot be associated with any solutions of
the Sine-Gordon equation, i.e. all such iterated solutions are solutions of
reduced equation (\ref{equa}). The above formulas just allow to connect an
infinite set of solutions determined by different expressions $h(w)$.

\section{Conclusion}

It is well known that integrable systems conditionally can be split on two
wide classes: $S$ and $C$ integrable systems. Usually, we understand that $C$
integrable systems are linearizable systems by appropriate transformations.
General solutions of $C$ integrable systems can be expressed explicitly, and
these solutions are parameterized by arbitrary functions. $S$ integrable
systems possess infinitely many particular solutions which are parameterized
by sufficiently many arbitrary constants in general. Also, we know that some 
$S$ integrable systems can be degenerated (for instance, the Sinh-Gordon
equation $u_{xt}=c_{1}e^{u}+c_{2}e^{-u}$ in a particular case becomes the
Liouville equation $u_{xt}=e^{u}$, whose general solution is parameterized
by two arbitrary functions of a single variable; see other detail in \cite%
{GZ}). In this paper we considered Constant Astigmatism equation (\ref%
{complex}), which is integrable by the Inverse Scattering Transform (see
again detail in \cite{MM}, \cite{MM1}, \cite{MM2}, \cite{MM3}, \cite{PZ})
but also admits particular solutions parameterized by an arbitrary function
of a single variable even in a nondegenerate case. We hope that particular
solutions parameterized by arbitrary functions of a single variable for some
other important integrable systems will be found soon.

In the previous Section we consider transformations for the Constant
Astigmatism equation, which replicate solutions of its reduced version (\ref%
{equa}) only. We hope that a more general transformation connecting
solutions of (\ref{equa}) and (\ref{complex}) can be found.

\section*{Acknowledgement}

We thank Michal Marvan for fruitful discussions. MVP was supported by GA\v{C}%
R under the project P201/11/0356, by the RF Government grant
\#2010-220-01-077, ag. \#11.G34.31.0005, by the grant of Presidium of RAS
\textquotedblleft Fundamental Problems of Nonlinear
Dynamics\textquotedblright\ and by the RFBR grant 11-01-00197. MVP also
would like to thank the National Group for Mathematical Physics (GNFM) for
funding his visit at the Department of Mathematics and Informatics of
University of Messina.

\end{document}